# Night sky brightness above Zagreb

## 2012.-2017.

**Željko Andreić[1]**

University of Zagreb, Faculty of Mining, Geology and Petroleum Engineering,

Pierottijeva 6, 10000 Zagreb, Croatia

**Abstract**

The night sky brightness at the RGN site (near the centre of Zagreb, Croatia) was monitored form January 2012. to December 2017. The gathered data show that the average night sky brightness in this period did not change significantly, apart from differences caused by yearly variations in meteorological parameters. The nightly minima, maxima and mean values of the sky brightness do change considerably due to changes in meteorological conditions, often being between 2 and 3 magnitudes. The seasonal probability curves and histograms are constructed and are used to obtain additional information on the light pollution at the RGN site. They reveal that the night sky brightness clutters around two peaks, at about 15.0 mag/arcsec$^2$ and at about 18.2 mag/arcsec$^2$. The tendency to slightly lower brightness values in spring and summer can also be seen in the data. Two peaks correspond to cloudy and clear nights respectively, the difference in brightness between them being about 3 magnitudes. A crude clear/cloudy criterion can be defined too: the minimum between two peaks is around 16.7 mag/arcsec$^2$. The brightness values smaller than this are attributed to clear nights and vice-versa. Comparison with Vienna and Hong-Kong indicates that the light pollution of Zagreb is a few times larger.

**Keywords**

Light pollution, night sky brightness, site testing, atmospheric effects

## 1. Introduction

The light pollution is most simply defined as any artificial light that spills into the environment. More elaborate definitions of various aspects of light pollution can be found in literature (see for instance **Mizon 2012**) or on the web pages devoted to light pollution, like those of the International Dark Sky Association (**IDA1 2017**) or Wikipedia (**Wiki2 2017**).

The impact of the light pollution on the environment and humans is very complex and still not well understood (see for example **Narisada 2004**). In all studies of such an impact the most important parameter is the amount (intensity) of the light pollution and its duration. Techniques of measuring and characterization of light pollution are still evolving and just a few instruments and procedures do exist today (**Hanel 2017**). Apart from global data on light pollution in form of various satellite images and maps (see for instance **Falchi 2016**), data for particular sites or environments are still scarce and not monitored on a regular basis. In Croatia, the light pollution was, to our knowledge, for the first time mentioned in an article in a popular astronomy magazine in 1993 (**Andreic 1993**). The first efforts to measure light pollution were taken during the 2002 Summer school of astronomy that took place in Višnjan, Istria (**ZEC 2018**). As at that time no dedicated instruments for measuring light pollution existed, the astronomical CCD camera was used in these attempts. In 2006, all-sky photography and SQM instruments were introduced and first usable data on LP were obtained, limited to several places in the Istria peninsula. The first model of light pollution in Croatia was finished in 2007. (**Andreic 2011**). The situation today is not much better. Apart from a few measurements that amateur astronomers did for their needs (**LPO1 2018**) and a few measuring campaigns in the past (see for instance **Andreic 2012, Sharma**

---

**2015**), the only systematic long-time observations of light pollution are those carried on at the RGN site (started in 2012.) and at the site of Merenje observatory, located about 30 km NW from Zagreb (started in 2014).

## 2. Data acquisition and reduction

The night sky brightness was measured by an SQM-LE instrument (**Unihedron1 2017**) permanently placed on the roof of the building of the Faculty of Mining, Geology and Petroleum Engineering in Zagreb (45.80701º N, 15.96398º E, approx. 150 m above sea level). The building itself is near the town centre, about 1.2 km air-line from the Zagreb main square. The instrument looks straight into the zenith. As the faculty building is among the highest in the surrounding area, there is no direct influence of lights from nearby buildings or street lighting on the measurements. The data are read by a remote PC connected to the SQM-LE by an ethernet cable. The instrument operates continuously, apart from power and hardware failures that produced several large and a lot of smaller "holes" in the data set in the abovementioned period.

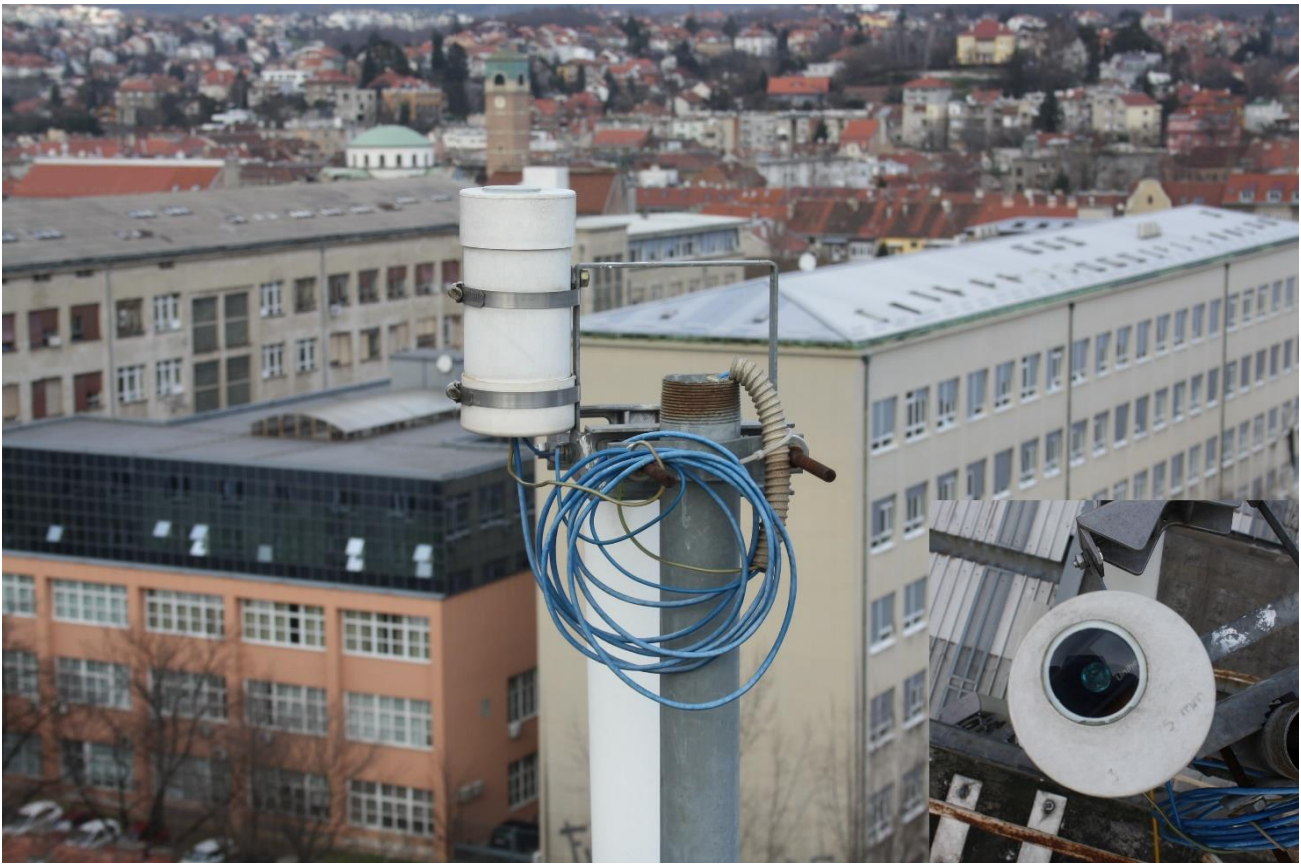

**Figure 1:** The SQM-LE instrument is mounted on the roof of the faculty building. The instrument itself is put inside a protective box (white) for protection from environmental impacts. The only maintenance required is periodical cleaning of the glass window on the top of the protective box. The inset at bottom-right shows top view of the enclosure, with glass window and the instrument visible inside.

The SQM measures the sky brightness in standard astronomical units of magnitudes per square arc-second. This brightness scale is used through the paper. This astronomical brightness scale is a reversed logarithmic scale (**Wiki1 2017**) meaning that larger values on the scale represent smaller brightness. The measuring unit is called "magnitude" and is equal to the brightness ratio of

2.512.... (or, exactly 5th root of 100). If conversion to linear units is needed, one can use the following formula (**Unihedron2 2017**):

$$[\text{value in cd/m}^2] = 10.8 \times 10^4 \times 10^{(-0.4 \times [\text{value in mag/arcsec2}])}$$

The main reasons for the popularity of SQM instruments are affordability and ease of use. On the other hand, they are not built to professional standards and this should be kept in mind during the data analysis process. Both the manufacturer (**Unihedron3 2017**) and the independent analyses of the instrument accuracy (**Cinzano 2005, Schnitt 2013**) lead to the same conclusion: the accuracy of the SQM is of the order of 10%, a very nice achievement when we take into consideration that this accuracy is kept over large range of brightness levels (about $10^4$ on linear scale).

The measured brightness values are displayed with two decimal places of the magnitude scale, indicating accuracy of the order of 1%. The second decimal place is not needed at all, as 10% accuracy is represented by tenths of the magnitude (i.e. the first decimal place), but the instrument shows it. In detailed analysis of our data we found out that internal A/D conversion in the SQM cannot produce all values of brightness levels that the display can show. The reason for such behaviour is hard to detect without complete knowledge about device construction and conversion process. This does not degrade the abovementioned 10% accuracy but can cause problems if one tries to do analysis on finer brightness levels, as some intensity values at 1% level are not present at all. This is nicely illustrated on the Fig. 2.

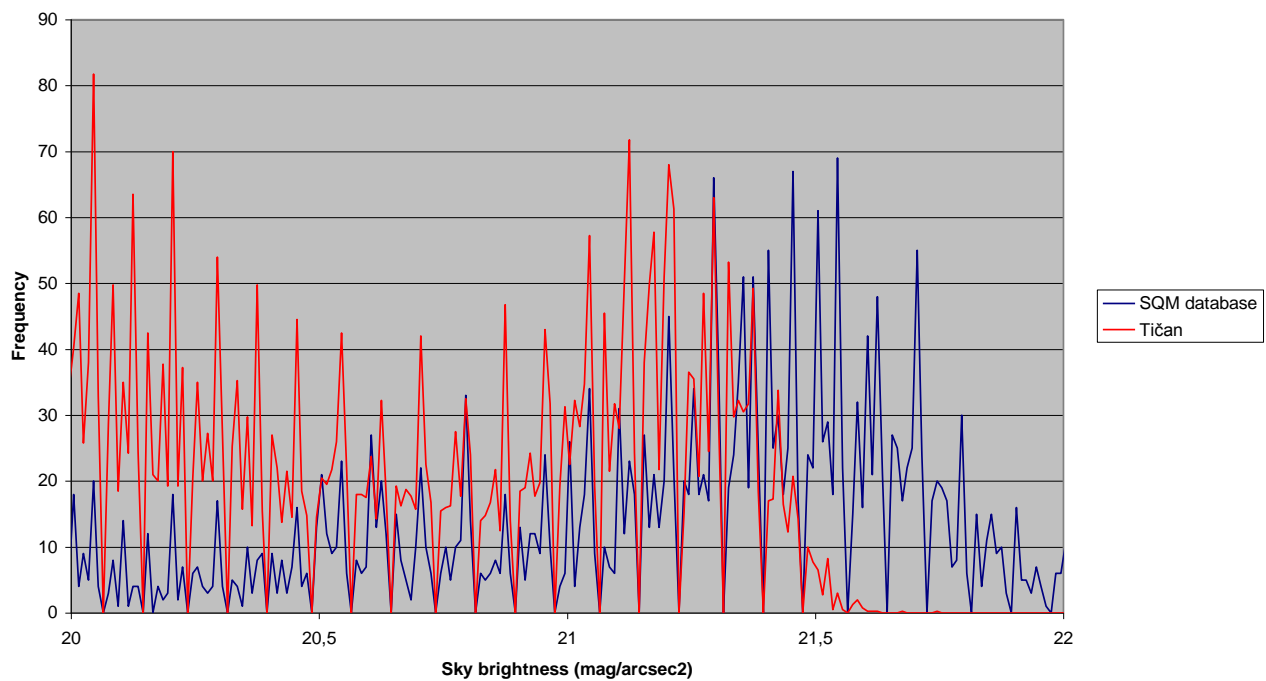

**Figure 2:** The detail of a histogram of measured values of the night sky brightness with 0.01 magnitude bin width clearly shows that some brightness values never appear in the SQM measurements. At the same time nearby values (a few hundredths of magnitude larger or smaller) are quite common, giving histogram the appearance of a periodically oscillating curve. The red line represents histogram created from 32000+ measurements made with an SQM-LE instrument at astronomical observatory Tičan (Istria peninsula, near the town of Višnjan), while blue line represents histogram created from 4000+ measurements from the world-wide base of SQM measurements (**SQM database 2017**) that is created by collecting data from observers all over the world. These observers used hand-held SQM or SQM-L devices of their own, meaning that almost each measurement was made

with a different device, proving that the effect is not malfunction of a particular unit but common characteristic of all SQM devices. To ease comparison, the Tičan curve frequencies are divided by 4.

The raw data (Fig. 3) were first reduced in size by rejecting measurements taken during daytime. The SQM cannot measure large sky brightness that appear during daytime, but is not smart enough to stop measuring during the day. Thus about 50% of the data are meaningless values that are removed from the data set before further analysis. After that, the data are divided into more manageable sets covering individual months of the year in question.

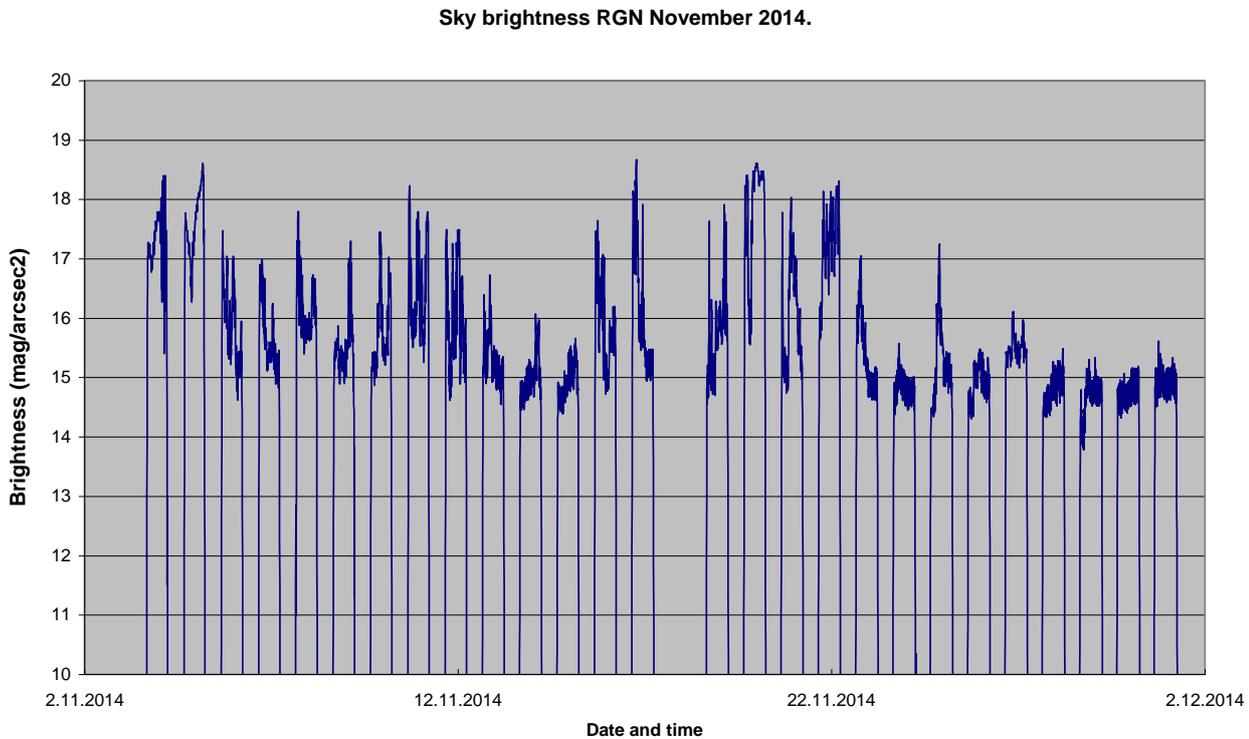

**Figure 3:** The raw measurements of the sky brightness, as taken by the SQM-LE instrument. Note that astronomical brightness scale is used. This is a reversed logarithmic scale (i.e. larger magnitude values represent lower brightness).

Next, the time stamps of individual data points are replaced by the ordinal number in the current data set. If this is not done, the daily gaps will still be present on the graphs. The intention behind the removal of the date/time stamps is to use graph area more efficiently. To retain basic information about the dates of the measurements in question, the day number (in the month in question) was kept in the dataset. An example plot of such a dataset is shown on the Fig. 4. Such plots were created for the whole-time period from January 2012. up to December 2017. These plots are very useful in interpreting the results of the measurements. They all can be accessed on the Light Pollution Observatory site (**LPO2 2018**).

The first question that arises when analysing such a dataset is weather the sky was clear, clouded or foggy. At least for a heavily light polluted site, as is this one, the answer can be found quite easily: the clouds scatter much more light downwards than the clear atmosphere, and consequently, the brightness of the night sky measured during cloudy nights is much larger, typically 3 or more magnitudes for this particular site. Also, the brightness oscillations are much faster and often larger than in the case of a clear night. The effect of the moonshine can be detected as slow, and rather smooth, rise (or drop) of the sky brightness. These conditions are illustrated on the Fig. 5. If needed, the periods when the Moon is above the horizon can be easily obtained from a planetarium or ephemerids software, for instance SkyChart (**SkyChart 2017**) or Ephem (**Ephem 2017**), both being freeware.

The way the analysis is carried on depends on the purpose for which the results are needed for. If we are looking for clear sky values, as needed by astronomers (both professional and amateur) we will search for the minimum brightness, as this determines the best observing conditions achievable at the site in question. Additionally, the frequency of occurrence of such conditions, and the duration of such periods of good sky conditions will be of interest.

On the other hand, if we are trying to assess the influence of light pollution on the biosphere, we will be more interested in the maximal sky brightness, as this is supposed to cause the most impact on biosphere, and in the frequency of occurrence and the duration of such conditions. The mean values of sky brightness on nightly, monthly and yearly basis could also be of importance. For such purposes the cloudiness and presence/absence of the Moon are of secondary interest, or not relevant at all.

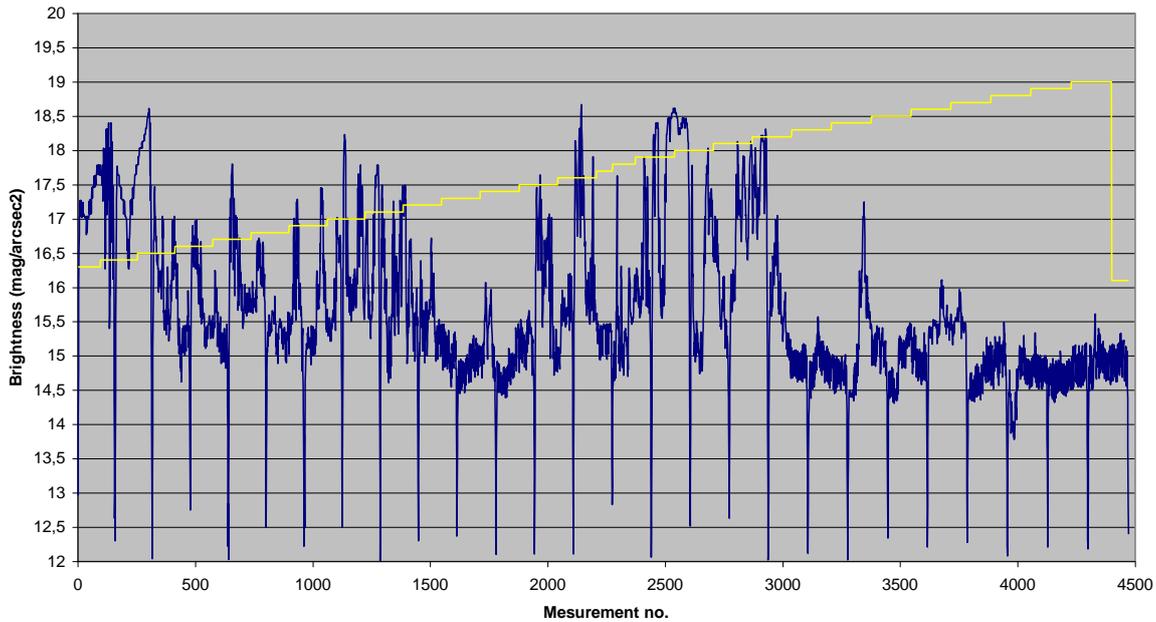

**Figure 4:** The sky brightness graph after removing day measurements and time stamps from dataset. The yellow line represents days in the current month and serves to detect any "holes" (missing data) in the dataset. It is created using the formula (day number)/10+16. Note that the day number changes at midnight.

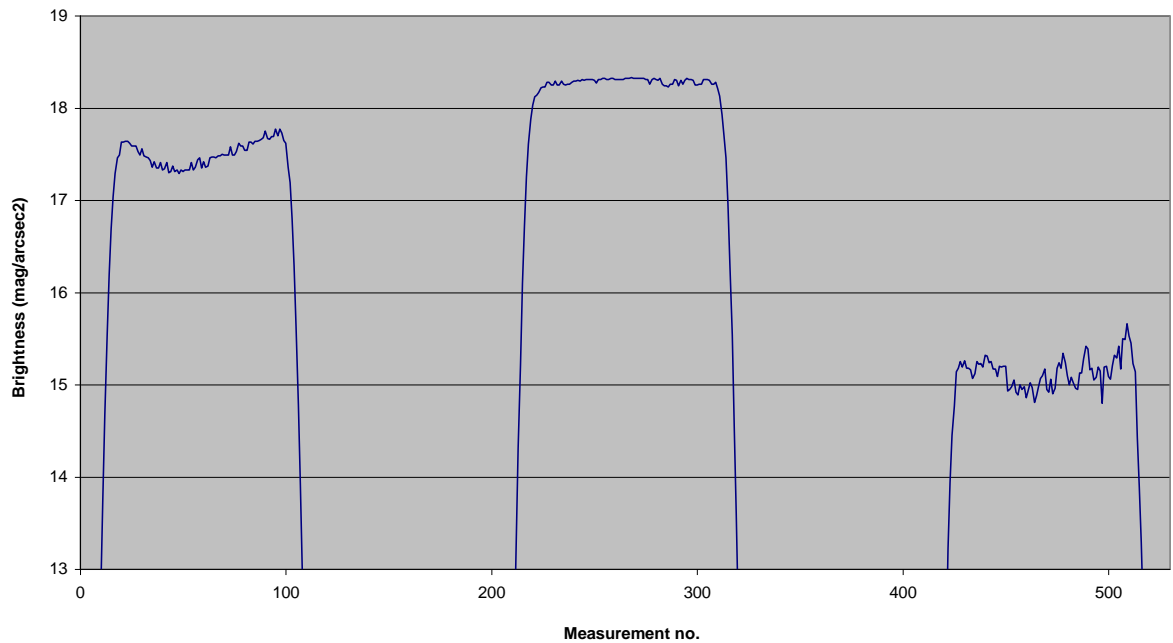

**Figure 5:** The examples of a clear night with a strong moonlight (left), a clear, moonless night (middle) and a cloudy night (right). All three examples are from measurements at RGN site. The effect of moonlight is more pronounced on sites with less light pollution.

To cover all these requirements, it was decided to determine minima, maxima and average values of sky brightness on a nightly basis. The results of these calculations are given graphically on the Figs. 6 to 8. After that, yearly statistics is extracted from the datasets and is summarized in the Table 1.

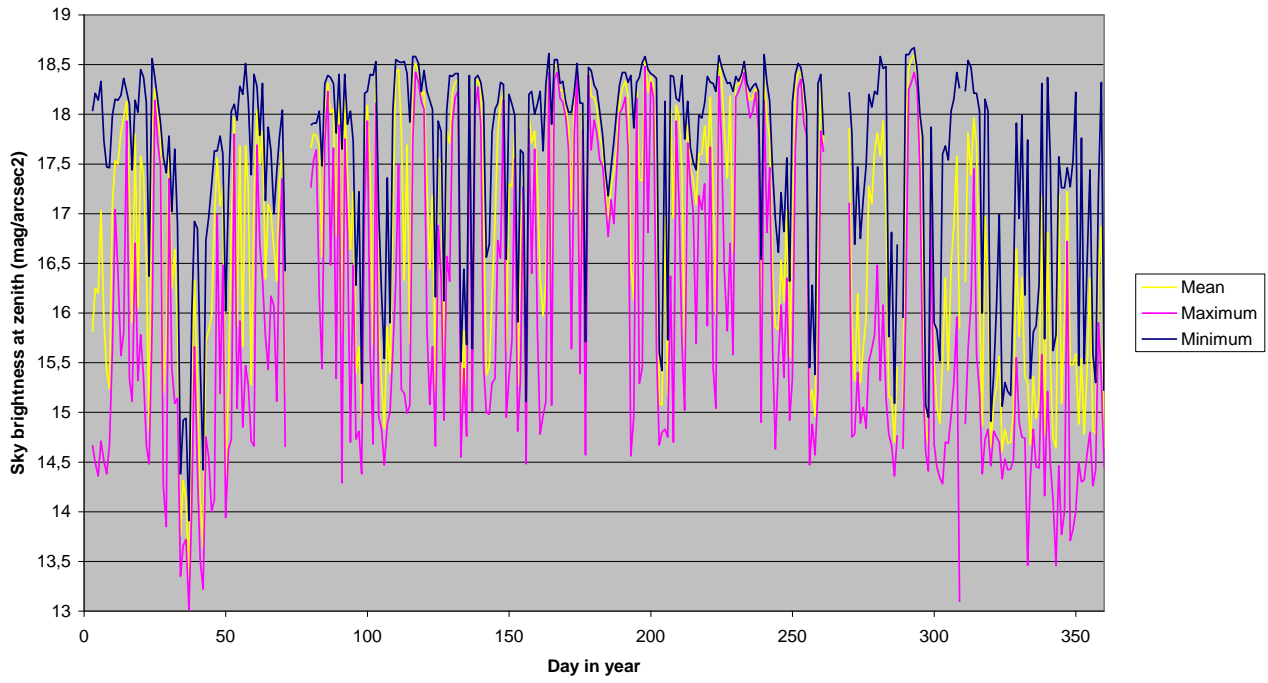

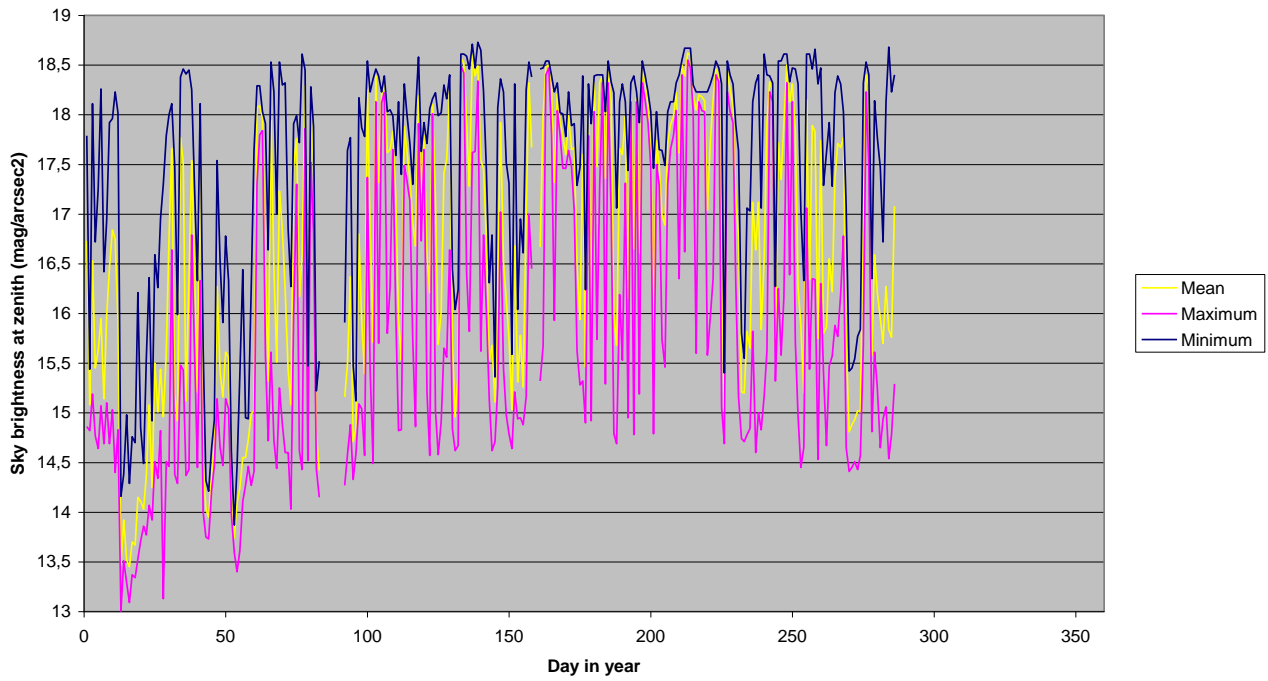

**Figure 6:** The nightly minima, maxima and average values of sky brightness for years 2012. and 2013.

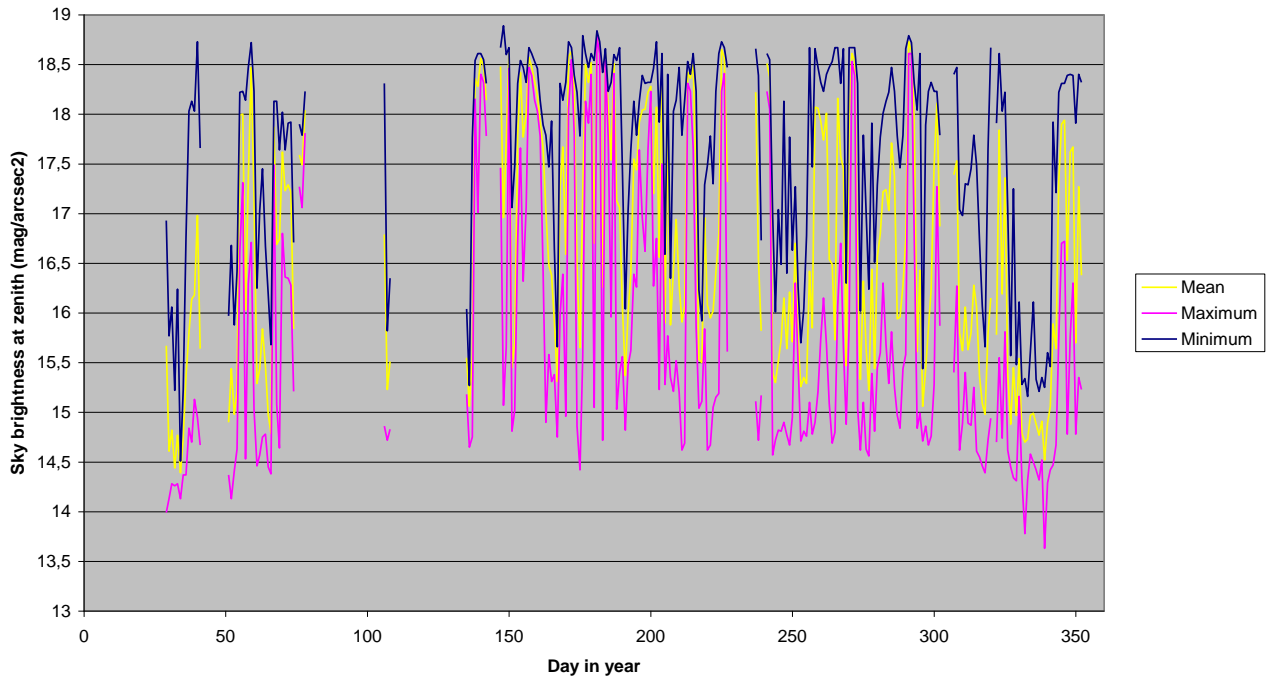

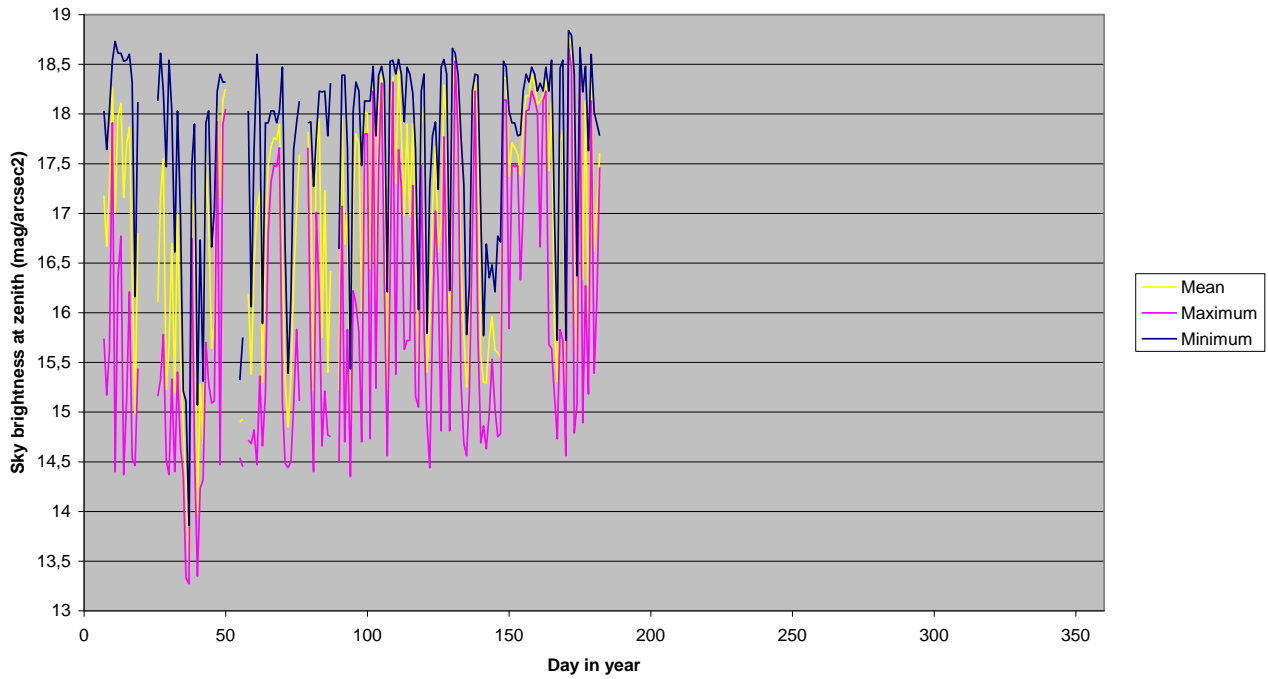

**Figure 7:** The nightly minima, maxima and average values of sky brightness for years 2014. and 2015.

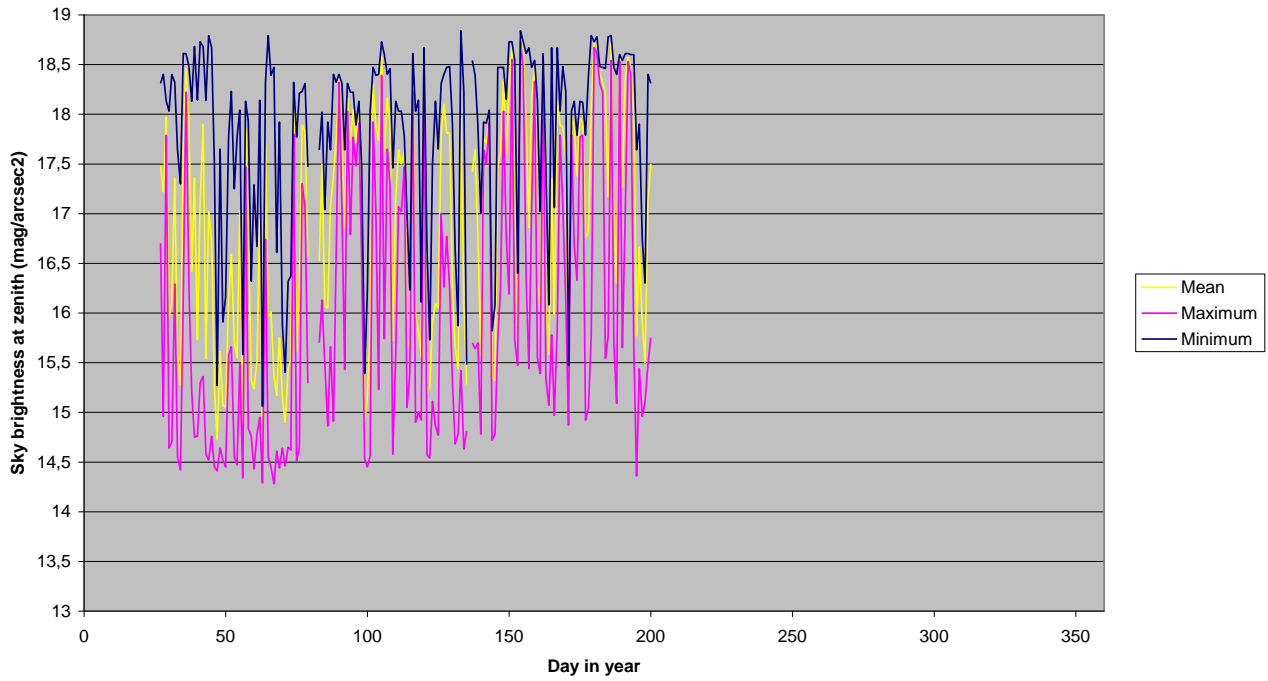

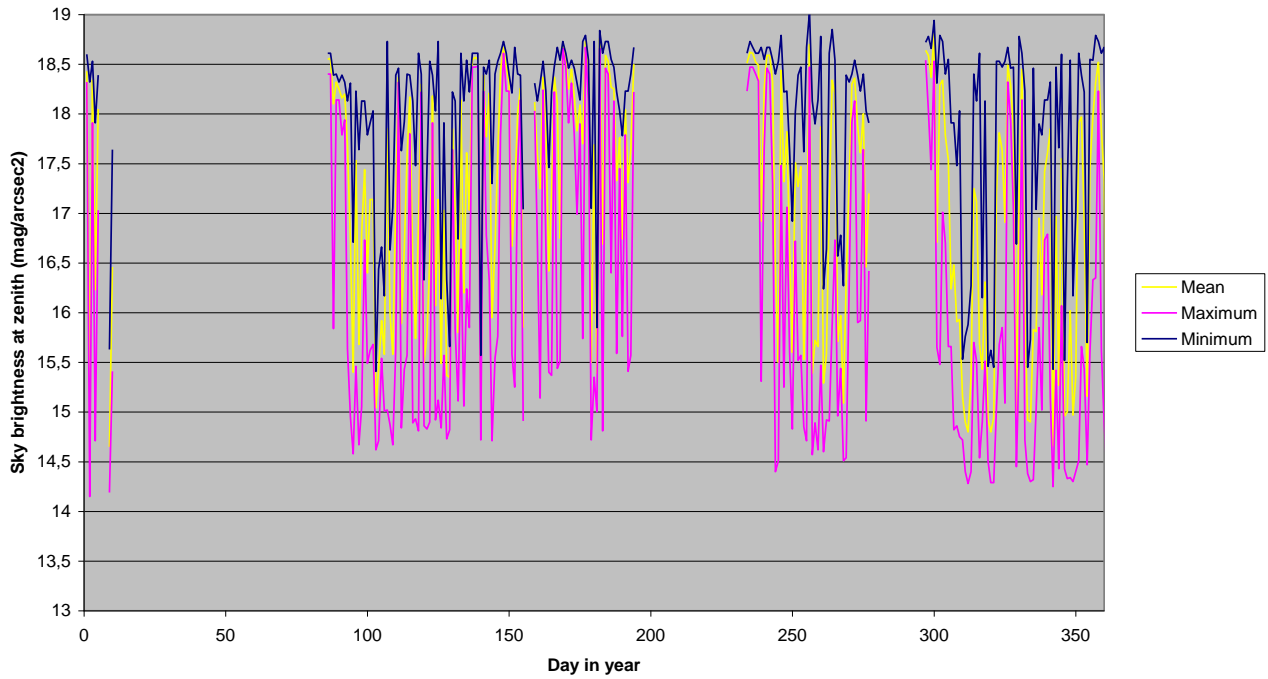

**Figure 8:** The nightly minima, maxima and average values of sky brightness for years 2016. and 2017.

| Year | Sky brightness | | | nightly variations | | | number of nights | missing nights |
|---|---|---|---|---|---|---|---|---|
| | min. | max. | mean | min. | max. | mean | | |
| 2012. | 13.01 | 18.67 | 16.80 | 0.08 | 5.17 | 1.66 | 346 | 20 |
| 2013. | 13.00 | 18.73 | 16.64 | 0.07 | 4.18 | 1.70 | 276 | 89 |
| 2014. | 13.63 | 18.89 | 16.66 | 0.06 | 3.87 | 1.94 | 244 | 121 |
| 2015. | 13.27 | 18.84 | 16.90 | 0.08 | 4.33 | 1.78 | 161 | 204 |
| 2016. | 14.20 | 18.94 | 16.99 | 0.06 | 4.27 | 1.83 | 324 | 41 |
| 2017. | 14.15 | 19.03 | 17.08 | 0.07 | 4.20 | 1.88 | 227 | 138 |
| mean | 13.54 | 18.85 | 16.85 | 0.07 | 4.34 | 1.80 | 263 | 102 |

**Table 1:** Yearly statistics of nightly minima, maxima and average values of sky brightness and its nightly changes (variations) for years 2012. to 2017. All brightness values are expressed in magnitudes per arc-second squared. The last row gives mean values for the whole period of measurements (2012.-2017.)

## 3. Results and discussion

In the period between January 2012. and December 2017. about 1/4 (28%) of the data was lost for different reasons. The most problematic in this aspect are years 2015. and 2016., for which only data for the first half of the year exist. However, the remaining data are more than enough for a sound analysis, and summary statistics (see Table 1.) does not show any significant differences from year to year that could be related to the missing data.

The first fact that strikes the eye when looking at the graphs on Figs. 6 to 8 is that all parameters (i.e. minima, maxima and mean values of the sky brightness) change considerably on nightly basis. The main cause for these variations are changes in meteorological conditions (clouds, fog, atmospheric transparency) that reflect themselves in amount of light pollution caused. The primary source of light pollution (artificial light) does not change so much and not so rapidly, a fact that can be confirmed by the long-term stability of average night sky brightness over the years (see Table 1).

Also, rapid changes that happen during one night are considerable, often being grater that 2 or even 3 magnitudes. Even the mean variations are considerable, being between 1.7 and 1.9 magnitudes from year to year (expressed as yearly averages).

Due to these rapid variations, Figs. 6 to 8 look crowded and difficult to interpret. Thus, the most important statistical data about light pollution on the RGN site are gathered in the Table 1, and later also presented in different form on Figs. 9 and 10.

The Table 1 summarises yearly minimums, maximums and average values of the measured sky brightness followed by the nightly variation of these values (also expressed as minimal, maximal and mean values observed over the year), together with information on number of nights that provided the data for the statistics. The main conclusion that can be drawn from the Table 1 is that the average level of light pollution at the RGN site does not change significantly during the monitoring period (2012. to 2017.). This does not mean that the light pollution produced by the town of Zagreb is also stable, it just states the fact that it is stable near the centre of the town. Considering that the centre is fully formed and illuminated a long time ago, it is reasonable to expect that no major changes in amount of public lighting (being the main cause of the light pollution in a town) occurs here. However, the situation might be quite different in suburbs that are rapidly growing, and the public lighting network expands there. What changes this growth produces cannot be concluded from measurements taken near the town centre (the RGN site) and would require additional measuring sites in suburbs that currently do not exist.

Apart from the basic information collected in the Table 1, the probability that at any given moment (at night of course!) the sky brightness will be smaller (or greater) than a certain value can be needed. This data is given on the Fig. 9 in form of cumulative probability curves derived from the complete dataset at hand. One must keep in mind that seasonal variations are quite large from year to year, so these curves can be used as a rough guide only. Also, it should be noted here that meteorological seasons are used through this text.

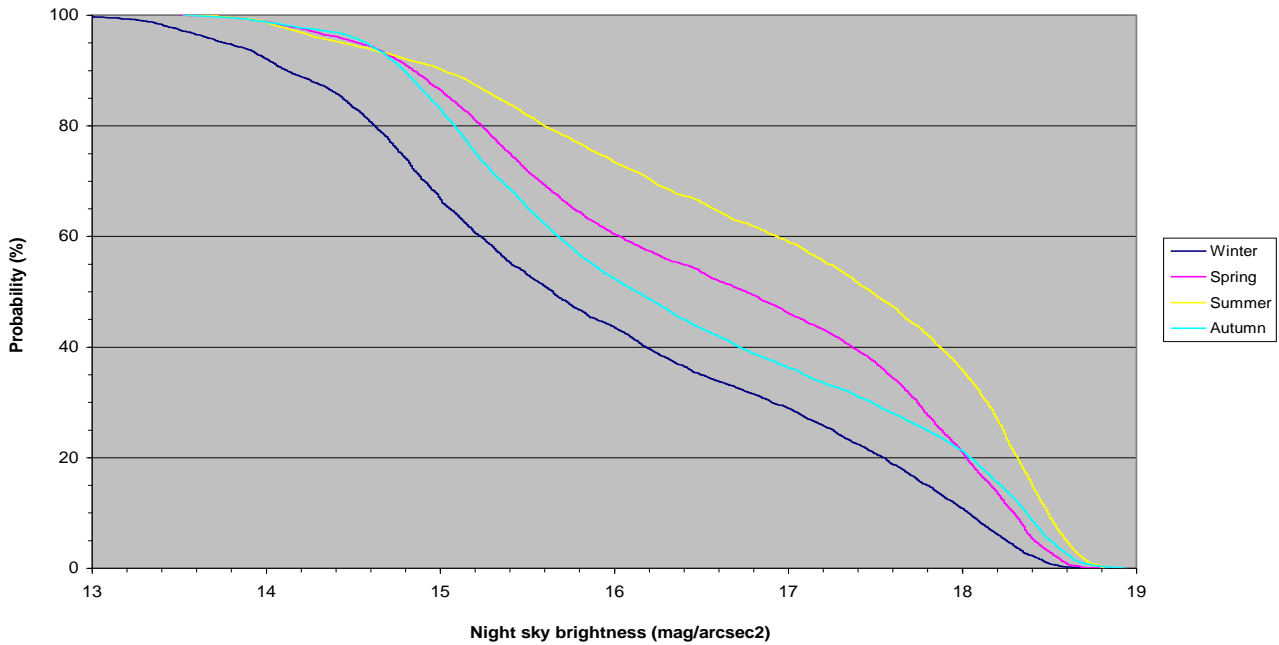

**Figure 9:** The average cumulative probability that the night sky brightness will be smaller than a given value, derived from all measurements available during the 2012.-2017. period.

Finally, the histogram of sky brightness is constructed from all the available data, also on seasonal basis. It gives the probability that at any given moment the sky brightness will have a certain value (or more precisely, will be in the corresponding brightness bin). The histogram bins are 0.1 mag in width, corresponding to the accuracy of the SQM device and avoiding bad data representation at finer scales, described before. Again, the seasonal variations in data from year to year are considerable, so the histogram should be used with some caution in mind.

Both Fig. 9 and Fig. 10 confirm the fact that summer (and spring) provide better observing conditions (less light pollution) than autumn and winter. One should however also consider that duration of the night changes considerably over the year, and that in winter/autumn nights quite often start as clear and end as clouded or fogged.

The histogram on the Fig. 10 reveals an interesting fact: the brightness values clutter around two peaks, one at about 15.0 mag/arcsec$^2$ and the other at about 18.2 mag/arcsec$^2$, both showing tendency to move slightly toward lower brightness values in spring and summer. The explanation of this effect is rather straightforward: the nights are either mostly clear or mostly cloudy. Clear nights result in lower sky brightness, the values clustering around the second peak, and the cloudy nights result in much larger sky brightness that clusters around the first peak. The difference between them is about 3 magnitudes, in accordance with previous conclusions. The slight drift toward lower brightness in spring/summer is result of generally dryer and more transparent atmosphere in this period of the year. Note that these facts cannot be read from the statistics gathered in the Table 1, as in doing this statistics all values are drawn in equally regardless of the sky condition. We can derive a crude clear/cloudy criterion from the histogram: the minimum between two peaks is at around 16.7 mag/arcsec$^2$ so the brightness values smaller than that (larger magnitude values!) can be attributed to clear nights and vice-versa. This criterion is only approximate, but quite usable. Note also that positions of the two peaks and the minimum between them depend on the site in question, mostly on the strength of the light pollution, but also to some extent on meteorogical characteristics of the site in question. Because of that, this conclusion cannot be generalised without actual measurements on the site in question.

Last, but not least, the question arises how these data compare with other towns. Similar measurements were reported for Vienna in 2014 (**Puschnig 2014**) with mean night sky brightness of 16.3 and 19.1 mag/arcsec$^2$ for cloudy/clear conditions. The population of Vienna is around 1.7 million, while that of Zagreb is about 0.8 million. Yet, the cloudy night sky in Zagreb is 1.3 mag/arcsec$^2$ (about 3.2 times) brighter, while the clear night sky is about 0.9 mag/arcsec$^2$ (about 2.3 times) brighter, both values indicating that Zagreb is a lot more light-polluted than Vienna. More, the value obtained for the mean night sky brightness in Hong-Kong, a heavily populated metropolis with a population of about 7.1 million, is 16.8 mag/arcsec$^2$ (**Pun 2014**). The authors did a lot of efforts to exclude the influence of moonlight on the measurements but did not try to separate clear from cloudy conditions. The corresponding value for Zagreb is 16.9 mag/arcsec$^2$, indicating that in this case Zagreb is heavily over lighted too.

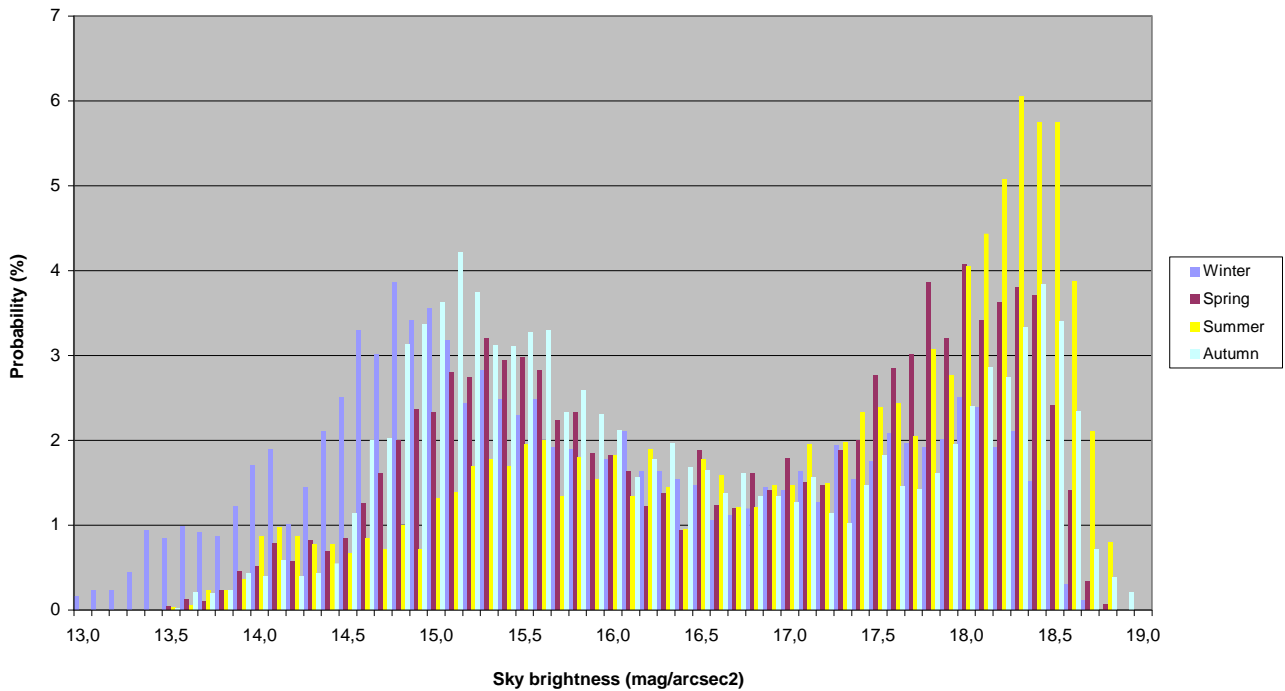

**Figure 10:** The histogram of the night sky brightness at RGN site derived from all measurements available during the 2012.-2017. period. Again, the histogram is created for meteorological seasons and should be taken as average of seasonal behaviour during the period 2012.-2017. The bins in the histogram are 0.1 magnitude in width.

## 4. Conclusions

The measurements of the night sky brightness at the RGN site cover the period between January 2012. and December 2017. with about 1/4 (28%) of the data missing. The statistical analysis of the data (see Table 1.) does not show any significant differences from year to year, apart from differences caused by yearly differences in meteorological conditions. It was found that all parameters (i.e. minima, maxima and mean values of the sky brightness) change considerably on nightly basis, mostly due to changes in meteorological conditions, often being larger than 2, sometimes even 3 magnitudes. The primary source of the light pollution (artificial light) does not change so rapidly, as is confirmed by long term stability of average night sky brightness over the measured period. Taking into account that the centre of Zagreb is fully formed a long time ago, this is expected. The situation might be quite different in suburbs that are rapidly growing but assessing the effects of this growth on the light pollution would require additional measuring sites in suburbs that currently do not exist.

The seasonal probability curves and histograms provide additional information on the light pollution at the RGN site. However, seasonal variations are quite large from year to year, so that information should be used with some caution in mind. The histograms reveal that the brightness values clutter around two peaks, one at about 15 mag/arcsec$^2$ and the other at about 18.2 mag/arcsec$^2$, with tendency to slightly lower brightness values in spring and summer. The two peaks correspond to cloudy and clear nights respectively, with the difference in brightness between them of about 3 magnitudes. The slightly lower brightness values observed in spring/summer are linked to dryer and more transparent atmosphere in this period of the year. A crude clear/cloudy criterion is derived too: the minimum between two peaks is at around 16.7 mag/arcsec$^2$. The brightness values smaller than that are attributed to clear nights and vice-versa. This conclusion is site dependent and cannot be generalised without measurements on other sites.

Last, but not least, comparison with similar measurements taken in Vienna and Hong-Kong indicate that Zagreb produces unproportionally large amounts of light pollution.

## 5. Acknowledgements

This work was partially supported by by the Ministry of Science and Sports of the Republic of Croatia scientific project 195-0000000-2233: "Erosion and landslides as geohazardous events (head Ž. Andreić) for priod 2011.-2014., by the University of Zagreb support "Mathematical research in Geology" (head T. Malvić) for 2016. and "Mathematical research in Geology II" (head T. Malvić) for 2017.